\documentclass{elsart1p}
\usepackage{graphicx,amssymb,epsfig,wrapft}
\begin{document}
\begin{frontmatter}
\title{Global analysis of hadron-production data in $e^+ e^-$ annihilation
       for determining fragmentation functions}

\author{M. Hirai$^{\, \rm a}$, S. Kumano$^{\, \rm b,c}$, 
        T.-H. Nagai$^{\, \rm c}$, M. Oka$^{\, \rm d}$,
        and K. Sudoh$^{\, \rm b}$}
\address{$^{\rm a}$ Department of Physics, Juntendo University, 
           Inba, Chiba 270-1695, Japan \\
         $^{\rm b}$ Institute of Particle and Nuclear Studies,
           High Energy Accelerator Research Organization (KEK) \\
           1-1, Ooho, Tsukuba, Ibaraki, 305-0801, Japan \\
         $^{\rm c}$ Department of Particle and Nuclear Studies,
           Graduate University for Advanced Studies \\
           1-1, Ooho, Tsukuba, Ibaraki, 305-0801, Japan \\
         $^{\rm d}$ Department of Physics, H-27,
           Tokyo Institute of Technology, Meguro, Tokyo, 152-8551, Japan}
\date{September 10, 2007}

\begin{abstract}
Fragmentation functions of pion, kaon, and nucleon are determined
by global analyses of hadron-production data in $e^+e^-$
annihilation. It is particularly important that uncertainties
of the fragmentation functions are estimated for the first time.
We found that light-quark and gluon fragmentation functions have
large uncertainties, so that one should be careful in using
these functions for hadron-production processes
in heavy-ion collisions and lepton scattering. The analysis is extended to 
possible exotic hadron search by fragmentation functions. We found that 
internal structure of $f_0 (980)$, such as $s\bar s$ or tetraquark
configuration, can be determined by noting differences between
favored and disfavored fragmentation functions. 
\end{abstract}

\end{frontmatter}

\section{Introduction}
\label{introduction}
\vspace{-0.2cm}

Fragmentation functions are becoming important recently because they
are used in describing hadron-production cross sections
in lepton scattering and heavy-ion collisions. 
There are a number of parametrization studies; however, obtained
functions are much different depending on analysis groups.
It suggests that the functions should not be well determined
at this stage although accurate determination could be crucial for
discussing nucleon spin and quark-hadron matters. Considering
this situation, we determined uncertainties of the fragmentation
functions in Ref. \cite{hkns07}.
The uncertainties are obtained in both leading order (LO) and
next-to-leading order (NLO).
This study is then extended to propose that the fragmentation
functions can be used for searching exotic hadrons by noting
differences between favored and disfavored functions \cite{hkos07}.
In the following, these studies are explained.

\section{Fragmentation functions and their uncertainties 
         for $\pi$, $K$, and $p$}
\label{ffs}
\vspace{-0.2cm}

The fragmentation function for a hadron $h$ is defined by
$F^h(z,Q^2) =  d\sigma (e^+e^- \rightarrow hX)/dz/\sigma_{tot}$,
where $d\sigma (e^+e^- \rightarrow hX)/dz$ is the hadron
production cross section and $\sigma_{tot}$ is
the total hadronic one. The variable  
$Q^2$ is given by the center-of-mass energy squared ($Q^2=s$),
and $z$ is defined by $z = E_h/(\sqrt{s}/2)$ with the hadron energy $E_h$.
The fragmentation should be described by the sum of partonic contributions
$F^h(z,Q^2) = \sum_i \int^{1}_{z} dy \, C_i(y,\alpha_s)  D_i^h (z/y,Q^2)/y$, 
where $C_i(z,\alpha_s)$ is a coefficient function and
$D_i^h(z,Q^2)$ is the fragmentation function
of the hadron $h$ from a parton $i$. 
The fragmentation functions are usually parametrized in the form:
$D_i^h(z,Q_0^2) = N_i^h z^{\alpha_i^h} (1-z)^{\beta_i^h}$, 
where $N_i^h$, $\alpha_i^h$, and $\beta_i^h$ are the parameters
which are determined by a $\chi^2$ analysis of
the $e^+e^- \rightarrow hX$ data.
Uncertainties of the determined functions are calculated by the Hessian
method.

\noindent
\begin{figure}[b]
\parbox[t]{0.48\textwidth}{
   \begin{center}
       \epsfig{file=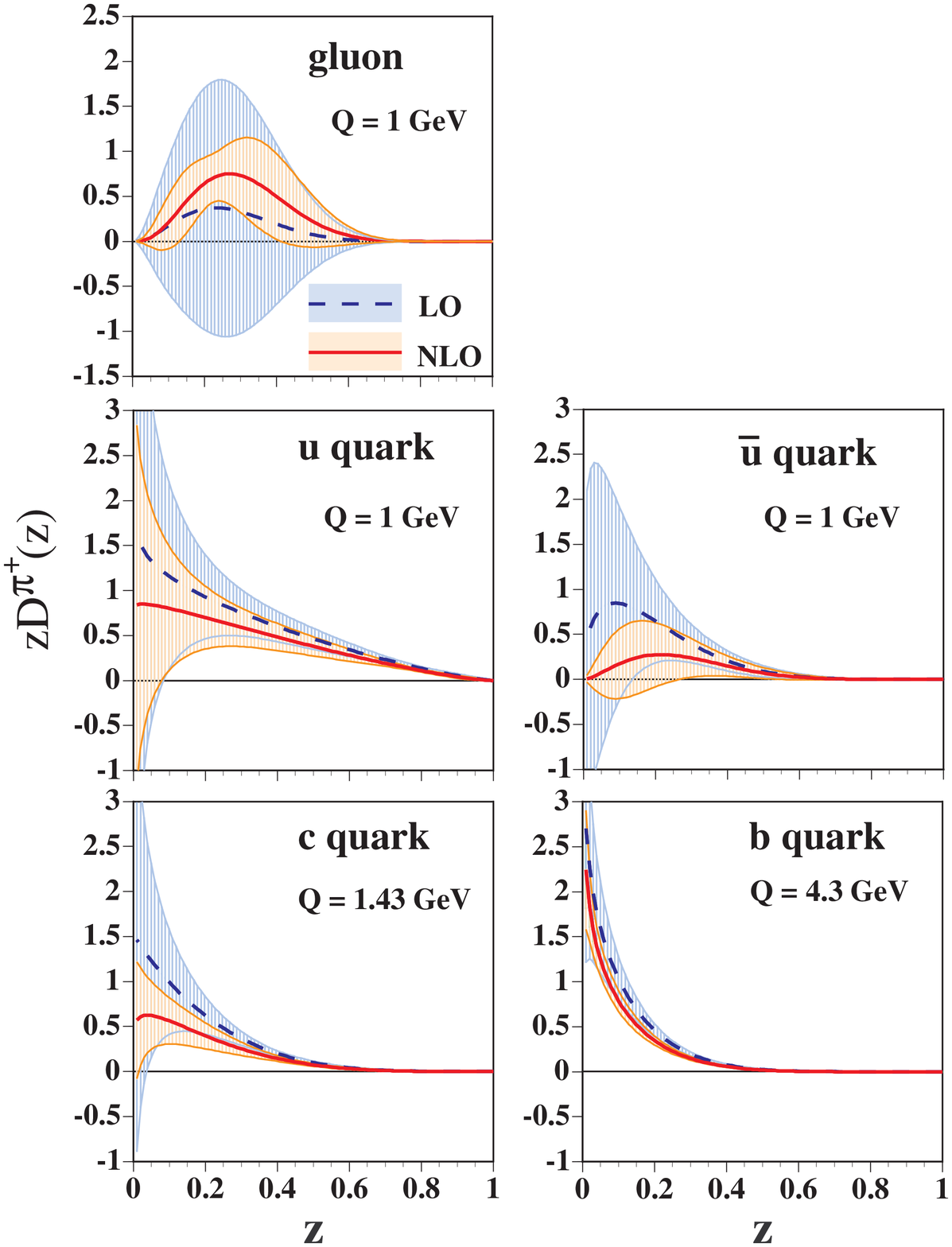,width=5.0cm} \\
       \vspace{-0.2cm}
       \caption{Fragmentation functions for $\pi^+$ and their
         uncertainties are shown in the LO and NLO.}
       \label{fig:pion-ff-q-1}
   \end{center}
}\hfill
\parbox[t]{0.48\textwidth}{
   \begin{center}
       \vspace{-0.1cm}
       \epsfig{file=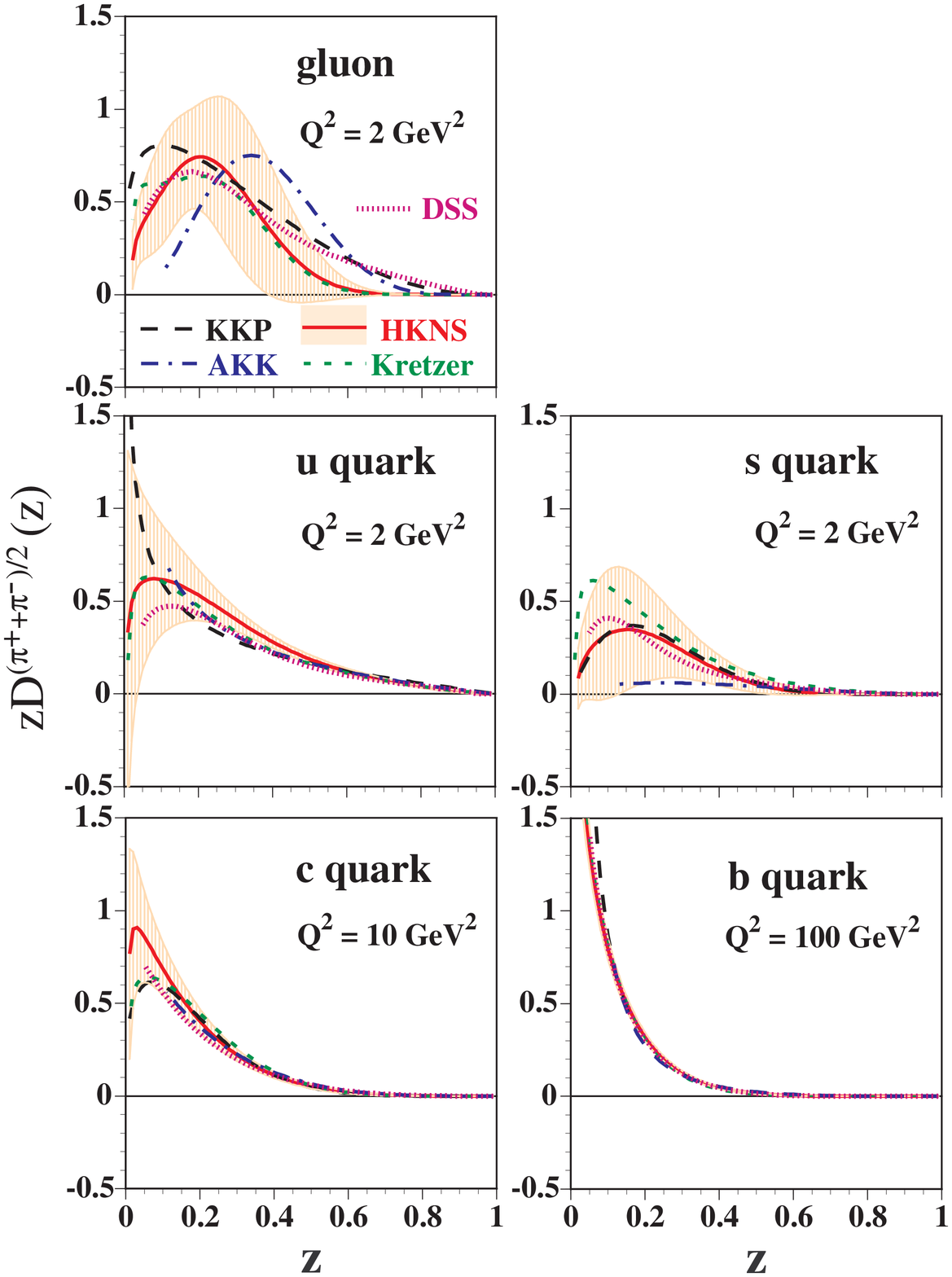,width=5.0cm} \\
       \vspace{-0.2cm}
       \caption{Fragmentation functions for $(\pi^+ +\pi^-)/2$
         are compared with other NLO analysis results.}
       \label{fig:pi-comp-ffs07}
   \end{center}
}
\end{figure}
\vspace{-0.40cm}

Determined LO and NLO fragmentation functions and their uncertainties
are shown in Fig. \ref{fig:pion-ff-q-1} for $\pi^+$ at $Q^2$=1 GeV$^2$,
$m_c^2$, or $m_b^2$ \cite{hkns07,web}.
The uncertainties indicate that the functions
are determined more accurately in the NLO in comparison with the LO.
However, the uncertainties are large even in the NLO analysis 
particularly in the gluon and light-quark functions. 
Since these functions are used for investigating the origin of
nucleon spin and properties of quark-hadron matters in
hadron-production processes, the uncertainties
should be taken into account in drawing any conclusion from
hadron-production data. Our codes for calculating the determined
functions and their uncertainties can be obtained from our web site
\cite{web}.

Next, the obtained functions denoted as HKNS (Hirai, Kumano, Nagai, Sudoh)
\cite{hkns07} are compared with other analysis results by
KKP (Kniehl, Kramer, P\"otter), Kretzer, AKK (Albino, Kniehl, Kramer), and 
DSS (de Florian, Sassot, Stratmann) for the pion ($(\pi^++\pi^-)/2$)
in Fig. \ref{fig:pi-comp-ffs07} at $Q^2$=2, 10, or 100 GeV$^2$.
Although there are huge differences between the parametrizations
in the gluon and strange-quark functions, all the curves are roughly
within our uncertainty bands. It means that all the analyses are 
consistent with each other in spite of the large differences. 

\section{Exotic hadron search by fragmentation functions}
\label{ffs}
\vspace{-0.2cm}

We applied our analysis to exotic hadron search. 
There are two types in the fragmentation functions:
favored and disfavored ones.
The favored fragmentation means a fragmentation from a quark
or an antiquark which exists in a hadron as a constituent
in a quark model, and the disfavored means a fragmentation
from a sea quark. Finding differences between these function,
we should be able to find internal structure of hadrons \cite{hkos07}.
As an example, we investigated $f_0 (980)$, whose structure
has been controversial. In a naive quark model, it may be
interpreted as $(u\bar u+d\bar d)/\sqrt{2}$; however, it
contradicts with experimental data of strong-decay 
and $\gamma\gamma$ widths. According to lattice QCD analysis,
its mass is too small to be interpreted as a glueball state.
Therefore, remaining possibilities are $s\bar s$ and
tetraquark $(u\bar u s\bar s+d\bar d s\bar s)/\sqrt{2}$ 
(or $K\bar K$) states.

\vspace{0.0cm}
\begin{wrapfigure}{r}{0.32\textwidth}
   \vspace{-0.25cm}
   \begin{center}
       \epsfig{file=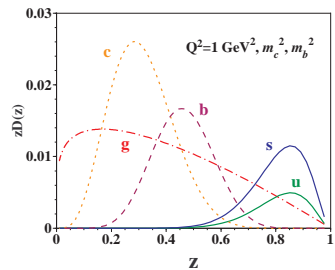,width=0.32\textwidth} \\
       \vspace{-0.3cm}
\caption{Determined fragmentation functions of $f_0(980)$.}
\label{fig:ff-f0}
   \end{center}
   \vspace{-0.3cm}
\end{wrapfigure}
\vspace{0.0cm}

\begin{table}[h]
\vspace{0.15cm}
\begin{tabular}{ccccc}
\hline
Type                   & Configuration 
                       & Status             
                       & Second moments
                       & \ \ \ Peak positions \      \\
\hline
Nonstrange $q\bar q$   & $(u\bar u+d\bar d)/\sqrt{2}$  
                       & Unlikely 
                       & $M_s<M_u<M_g$
                       & $z_u^{\rm max}>z_s^{\rm max}$   \\  
Strange    $q\bar q$   & $s\bar s$                 
                       & Possible
                       & $M_u  <   M_s \lesssim M_g$    
                       & $z_u^{\rm max}<z_s^{\rm max}$   \\  
\ Tetraquark (or $K\bar K$) \  
           & \ \ \ \ $(u\bar u s\bar s+d\bar d s\bar s)/\sqrt{2}$ \ \ \ \ 
           & \ \ \ \ Possible \ \ \ \  
           & \ \ \ \ $M_u \sim M_s \lesssim M_g$ \ \ \ \ 
           & \ \ \ $z_u^{\rm max} \sim z_s^{\rm max}$ \   \\  
Glueball               & $gg$ 
                       & Unlikely
                       & $M_u \sim M_s < M_g$
                       & $z_u^{\rm max} \sim z_s^{\rm max}$   \\ 
\hline
\end{tabular}
\vspace{0.10cm}
\caption{}
\vspace{-0.6cm}
{\ \ \ \ \ \ \ \ \ \ 
         . \ Possible $f_0(980)$ configurations and their features 
         in fragmentation functions at small $Q^2$.}
\label{tab:f0-config}
\end{table}

If the $f_0$ is an $s\bar s$ state,
the favored fragmentation from $s$ is possible if a gluon
is radiated from $s$, and then it splits into an $s\bar s$ pair
to form the $f_0$ meson. This process is of the order of $g^2$,
where $g$ is the coupling constant. The disfavored fragmentation
is proportional to $g^3$ by considering a gluon radiation
to have a color singlet $f_0$ state, so that its probability is
expected to be smaller than the favored one ($M_u<M_s$ in second
moments). In this way, we obtain
characteristic features in second moments ($M_i$) of the fragmentation
functions and in their functional peak positions ($z_i^{\rm max}$) at
small $Q^2$ ($\sim$1 GeV$^2$) in Table \ref{tab:f0-config}
\cite{hkos07}.

\vspace{+0.03cm}
We show actual analysis results for the $f_0$ fragmentation functions
in Fig. \ref{fig:ff-f0}. The $s$- and $u$-quark functions indicate
valence-like structure which is peaked at large $z$.
This fact suggests a tetraquark configuration. However, the relation
$M_u < M_s$ indicates an $s\bar s$ type configuration.
These conflicting results are obtained mainly because
the functions are not accurately determined from the current $e^+e^-$
data, although there is a possibility of an admixture state.
If uncertainties of the functions are estimated, they are ten times
larger than the functions themselves \cite{hkos07}. However,
if accurate $e^+e^- \rightarrow f_0 X$ data are obtained,
internal structure should be determined.

\vspace{-0.3cm}


\end{document}